# A Kalman Filter Framework for Resolving 3D Displacement Field Time Series By Combining Multitrack Multitemporal InSAR and GNSS Horizontal Velocities

**Manoochehr Shirzaei**[1]
[1]Department of Geosciences, Virginia Tech, Blacksburg, VA, USA. USA (e-mail: shirzaei@vt.edu)

**ABSTRACT** The availability of Synthetic Aperture Radar (SAR) data from different sensors and observation of Global Navigation Satellite System (GNSS) has been growing worldwide. The complementary nature of InSAR and GNSS observations demands methodological advancements for integrating these datasets of variable accuracy, spatiotemporal sampling rate, and geometries to generate seamless maps of 3D time series that account for both observation's advantages. Here, I present an approach based on Kalman Filter, which recursively resolves the 3D displacement field time series by combining line-of-sight time series from at least one SAR viewing geometry and horizontal velocities from GNSS networks. I apply this method to 3 overlapping SAR frames in ascending and descending orbits of Envisat C-band and ascending orbit of ALOS L-band acquired over the San Francisco Bay Area from 2007 to 2011. The experimental results and validation tests against independent observations indicate that the presented approach can resolve 3D displacement field time series at mm-level accuracy comparable to GNSS accuracy but at 10s m spatial resolution.

## I. INTRODUCTION

Over the past three decades, Interferometric Synthetic Aperture Radar (InSAR) has caused significant advances in understanding Earth and environmental phenomena due to its unprecedented spatial resolution and global coverage [1, 2]. To overcome some of the limitations of conventional InSAR and mitigate impacts of environmental artifacts such as tropospheric delay and improve the signal-to-noise ratio via stacking, multitemporal InSAR approaches have been developed. These approaches investigate stacks of SAR images acquired over the same area with a similar viewing geometry [3-9]. The InSAR time series algorithms either apply filters to estimate and reduce the impact of atmospheric delay [3-5, 7] or exploit models informed by auxiliary information such as weather data [10-13].

InSAR measurements of the ground surface displacement are inherently one-dimensional (1D) in the line-of-sight (LOS) direction [14]. Thus, to obtain the 2D (east-west and up-down) or 3D (east-west, south-north, and up-down) or only up-down displacement fields, several approaches have been proposed to combine LOS observations from ascending and descending orbits and/or with other datasets (e.g., [15]). Examples include combining LOS measurements with (i) azimuth offset tracking [16-21], (ii) the direction of displacement field [22, 23], (iii) Global Navigation Satellite System (GNSS) observations [24-29], (iv) incorporating prior deformation models [30].

Furthermore, sophisticated tools such as the Kalman filter have been widely used to combine multisensor and multitrack datasets and create 3D displacement time series (e.g., [31-33]). To reliably resolve the 3D displacement field, particularly in the north-south direction, these approaches often assume a certain level of smoothness for the deformation field or integrate prior information such as strain models.

Here, I present a Kalman Filter framework, free from any prior information and models, to combine 1D LOS time series from different SAR viewing geometries with measurements of horizontal velocity fields obtained from GNSS stations and generate 3D displacement time series maps of surface displacement without imposing any assumption on the spatial or temporal behavior of the displacement field. I will apply this approach and validate the results using two overlapping frames of Envisat C-Band satellite in ascending and descending orbits and one frame of ALOS L-band satellite in ascending orbit, observing surface deformation over the San Francisco Bay Area from 2007 to 2011.

## II. Methods and Datasets

### A. Study area

The study area comprises San Francisco Bay Area (SFBA), in Northern California, with a population of ~7.7 million. The SFBA includes three major fault strands, Calaveras, Hayward, and San Andreas faults, which accumulate the majority of relative motion between the Pacific and the Sierra Nevada-Great Valley plates. These faults have slipped during seismic and aseismic events [34-36], and there is a high probability of a significant earthquake in the area within the next 30 years [37]. The SFBA also experiences land subsidence due to the compaction of aquifers [38] and sediments [39], exacerbating the relative sea-level rise and flooding hazards [26, 29].

### B. Datasets

To map the 3D surface displacement time series over the SFBA, I use three overlapping SAR frames acquired in ascending and descending orbits of Envisat C-band and ascending orbit of ALOS L-band satellites, spanning 2007/07/13 – 2010/10/17 (Figure 1 and Table I). Also, I use the GNSS horizontal velocities provided by Bürgmann et al. [39], a subset of the regional BAVU velocity field [40] that includes both continuous and campaign measurements. The GNSS dataset is randomly split into the tie and check datasets (Figure 1). The tie dataset (26 stations) is used to analyze InSAR datasets. While check dataset (28 stations) is set aside and only used for validating results. There are additionally 5 GNSS stations with continuous measurements during the observation period, which will be used for validating the obtained 3D displacement time series.

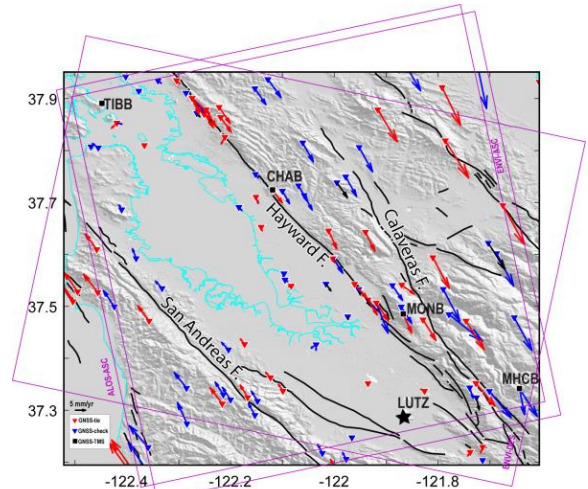

**FIGURE 1** study area and datasets, including footprint of SAR images and location of GNSS stations, used the analysis and validations. The Cayan line indicates the coastlines.

TABLE I

SAR DATA SET USED IN THIS STUDY, INCLUDING ENVISAT DATA ACQUIRED IN DESCENDING ORBIT TRACK 70 (INCIDENCE ANGLE = 23°, HEADING ANGLE = 193°) AND ASCENDING ORBIT TRACK 478 (INCIDENCE ANGLE = 23°, HEADING ANGLE = 350°), AS WELL AS ALOS SAR DATA OBTAINED IN ASCENDING ORBIT, FRAME 740 AND TRACK 222 (INCIDENCE ANGLE = 34.5°, HEADING ANGLE = 350°)

| Envisat-Des (YYYYMMDD) | Envisat-Asc (YYYYMMDD) | ALOS-Asc (YYYYMMDD) |
|---|---|---|
| 20070721 | 20070923 | 20070713 |
| 20070825 | 20071202 | 20070828 |
| 20070929 | 20080106 | 20071013 |
| 20071103 | 20080210 | 20071128 |
| 20071208 | 20080316 | 20080113 |
| 20080112 | 20080420 | 20080228 |
| 20080216 | 20080525 | 20080414 |
| 20080322 | 20080629 | 20080530 |
| 20080426 | 20080803 | 20080715 |
| 20080531 | 20080907 | 20081130 |
| 20080705 | 20081221 | 20090115 |
| 20080809 | 20090125 | 20090302 |
| 20080913 | 20090301 | 20090602 |
| 20081018 | 20090405 | 20090718 |
| 20081122 | 20090510 | 20091018 |
| 20090131 | 20090614 | 20100420 |
| 20090307 | 20090719 | 20100605 |
| 20090411 | 20090823 | 20100721 |
| 20090516 | 20090927 | 20101206 |
| 20090725 | 20091101 | |
| 20090829 | 20091206 | |
| 20091003 | 20100704 | |
| 20091107 | 20100808 | |
| 20091212 | 20101017 | |
| 20100116 | | |
| 20100220 | | |
| 20100327 | | |
| 20100501 | | |
| 20100605 | | |
| 20100710 | | |
| 20100814 | | |
| 20100918 | | |

## C. Multitemporal InSAR algorithm for LOS time series

To analyze each SAR dataset and create three precise maps of LOS displacement over SFBA, I implement the multitemporal algorithm detailed in [7, 41], which is a deviation of the small baseline subset approach [5]. In this approach, to calculate and remove the geometrical phase, following [14], I use satellite precise ephemeris data and a 10 m digital elevation model (DEM) provided by the Shuttle Radar Topography Mission [42]. To identify elite pixels, I use temporal coherence criteria [43], in which pixels that maintain a coherence above a certain threshold are selected. Next, I apply a minimum cost flow [44] algorithm to estimate absolute phase changes for each elite pixel [45]. I correct unwrapped interferograms for the effect of orbital error [46] and topography correlated atmospheric delay [47]. To this end, I apply filters that use the wavelet transform to identify and remove long-wavelength signal components and those that correlate with the DEM. The corrected and unwrapped interferograms are combined using a robust least-squares method[48] to create a time series of phase changes for each pixel. The time series are further refined using a wavelet-based adaptive filter [49] to reduce the effects of temporally uncorrelated atmospheric delay [7]. This filter identifies and then reduces the amplitude of the signals with temporal wavelength shorter than 300 days. The LOS velocities and associated standard deviation (STD) are shown in Figure 2. In this figure, I only show $n$ pixels with LOS observation in all three datasets, which will be used for the rest of our analysis.

## D. Multitrack multitemporal InSAR algorithm for 3D surface displacement time series

I further resample LOS time series over vector $t = \bigcup_s t_s$ to create 3 temporally overlapping datasets, where $t_s = \{t_1^s, t_2^s, \dots, t_{m_s}^s\}$, $s \in \{AlosAsc, EnviAsc, EnviDes\}$ and $m_s$ is number of acquisitions in each SAR dataset. Assuming $dl_s = \{dl_{t_1^s}, dl_{t_2^s}, \dots, dl_{t_{m_s}^s}\}$ and $\sigma_s^2 = \{\sigma^2_{t_1^s}, \sigma^2_{t_2^s}, \dots, \sigma^2_{t_{m_s}^s}\}$ are LOS observation and variance of dataset $s$, I apply a linear interpolation to obtain corresponding values a times $t$. Given the corresponding LOS observations and variances at times $a$ and $b$, the interpolated values, $\bar{dl}_c$ and $\bar{\sigma}_c^2$ at time $c$ are given [6];

$$\bar{dl}_c = \frac{t_c - t_a}{t_b - t_a}(dl_b - dl_a) + dl_a$$
$$\bar{\sigma}_c^2 = (\frac{t_c - t_a}{t_b - t_a}\sigma_b)^2 + (\frac{t_c - t_a}{t_b - t_a}\sigma_a)^2 + \sigma_a^2 \quad (1)$$

The interpolated observation vector $\bar{dl}_s = \{\bar{dl}_{t_1^s}, \bar{dl}_{t_2^s}, \dots, \bar{dl}_{t_m^s}\}$ and associated variances $\bar{\sigma}_s^2 = \{\bar{\sigma}^2_{t_1^s}, \bar{\sigma}^2_{t_2^s}, \dots, \bar{\sigma}^2_{t_m^s}\}$, where $m$ is the length of vector $t$.

Consider $q$ GNSS stations within the SAR frame whose horizontal velocities $[V_{x_i}, V_{y_i}]^T$ and variances $[\sigma^2_{x_i}, \sigma^2_{y_i}]^T$ are known, where $i = 1, \dots, q$. I further interpolate the GNSS velocities over InSAR pixels using a kriging algorithm to obtain a GNSS velocity per pixel $[\bar{V}_{x_i}, \bar{V}_{y_i}]^T$, where $i = 1, \dots, n$. The variances associated with the interpolated GNSS velocities at a pixel location $(\zeta_p, \eta_p)$ is obtained following;

$$\bar{\sigma}^2_{x_p} = \sigma^2_{x_i} \times \left(1 + \frac{D}{S}\right)^2$$
$$\bar{\sigma}^2_{y_p} = \sigma^2_{y_i} \times \left(1 + \frac{D}{S}\right)^2 \quad (2)$$

Where $D$ is the distance from the pixel $(\zeta_p, \eta_p)$ to nearest GNSS station in $km$ and $S$ is a scaling facor. Next, for each pixel at time $k$, I define 3 unknowns $[dx_{t_k}, dy_{t_k}, dz_{t_k}]^T$, which are the 3D displacement field vectors at a given time. To establish a Kalman Filter structure, the measurement model at a time $t_k$ is;

$$\bar{dl}_{t_k^{AlosAsc}} = C_x^{AlosAsc} \times dx_{t_k}$$
$$+ C_y^{AlosAsc} \times dy_{t_k}$$
$$+ C_z^{AlosAsc} \times dz_{t_k}$$
$$+ e_{t_k^{AlosAsc}}$$
$$\bar{dl}_{t_k^{EnvAsc}} = C_x^{EnvAsc} \times dx_{t_k}$$
$$+ C_y^{EnvAsc} \times dy_{t_k}$$
$$+ C_z^{EnvAsc} \times dz_{t_k} \quad (3)$$
$$+ e_{t_k^{EnvAsc}}$$
$$\bar{dl}_{t_k^{EnvDes}} = C_x^{EnvDes} \times dx_{t_k}$$
$$+ C_y^{EnvDes} \times dy_{t_k}$$
$$+ C_z^{EnvDes} \times dz_{t_k}$$
$$+ e_{t_k^{EnvDes}}$$

Where $e$'s are vectors of random noise and $C$'s are SAR satellite unit vectors projecting 3D displacement field onto LOS [50].

The system dynamic model is
$$dx_{t_k} = dx_{t_{k-1}} + (t_k - t_{k-1}) \times \bar{V}_x + w_{x_{t_{k-1}}}$$
$$dy_{t_k} = dy_{t_{k-1}} + (t_k - t_{k-1}) \times \bar{V}_y + w_{y_{t_{k-1}}} \quad (4)$$
$$dz_{t_k} = dz_{t_{k-1}} + w_{z_{t_{k-1}}}$$

Where $w's$ are vectors of random noise. Using Linear algebra, Equation 3 is expressed as

$$\boldsymbol{Y}_k = \boldsymbol{A}_k \boldsymbol{X}_k + \boldsymbol{e}_k$$

$$Y_k = [\overline{dl}_{t_k^{AlosAsc}} \quad \overline{dl}_{t_k^{EnvAsc}} \quad \overline{dl}_{t_k^{EnvDes}}]^T$$
$$X_k = [dx_{t_k} \quad dy_{t_k} \quad dz_{t_k}]^T$$
$$A_k = \begin{bmatrix} C_x^{AlosAsc} & C_y^{AlosAsc} & C_z^{AlosAsc} \\ C_x^{EnvAsc} & C_y^{EnvAsc} & C_z^{EnvAsc} \\ C_x^{EnvDes} & C_y^{EnvDes} & C_z^{EnvDes} \end{bmatrix}$$
$$e_k = [e_{t_k^{AlosAsc}} \quad e_{t_k^{EnvAsc}} \quad e_{t_k^{EnvDes}}]^T$$
(5)

recursive solution to the system of Equations 5 and 6 is given by [51, 52];

$$\widehat{X}_{k|k-1} = \widehat{X}_{k-1|k-1} + B_{k-1}u_{k-1}$$
$$P_{k|k-1} = P_{k-1|k-1} + Q_{k-1}$$
$$K_k = P_{k|k-1}A_k^T(A_k P_{k|k-1}A_k^T + R_k)^{-1}$$
$$\widehat{X}_{k|k} = \widehat{X}_{k|k-1} + K_k(y_k - A_k\widehat{X}_{k|k-1})$$
$$P_{k|k} = P_{k|k-1} - K_k A_k P_{k|k-1}$$
(7)

Where $\widehat{X}_{k|k}$ is the updated vector of state variables and $P_{k|k}$ is its updated variance-covariance matrix. This operation is performed for each pixel separately to create a time series of the 3D displacement field.

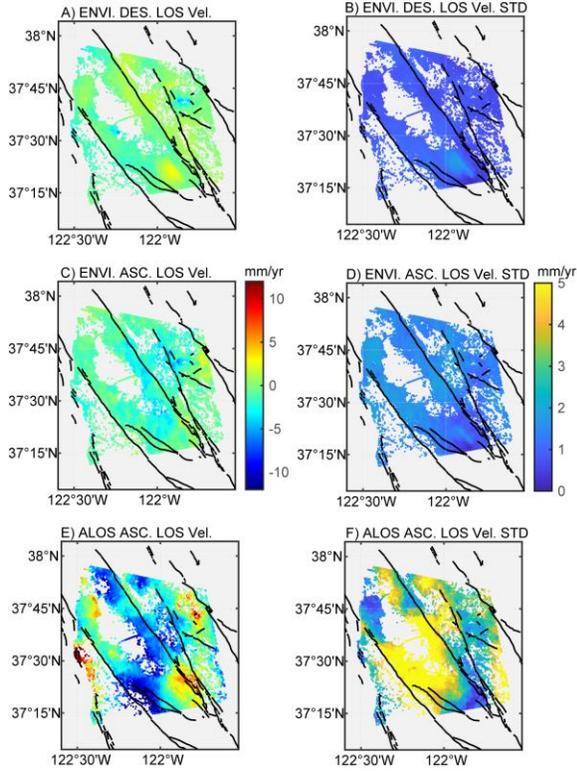

**FIGURE 2** LOS velocity and standard deviations (STD) associated with SAR each data set following InSAR time series analysis. Shown are collocated pixels that have observations in every dataset. A, B) Envisat ascending (incidence angle = 23o, heading angle = 350o), C, D) Envisat descending (incidence angle = 23o, heading angle = 193o) and E, F) ALOS ascending (incidence angle = 34.5o, heading angle = 350o. Black lines are major fault lines.

And Equation 4 is

$$X_k = X_{k-1} + B_{k-1}u_{k-1} + w_{k-1}$$
$$B_k = [(t_k - t_{k-1}) \quad (t_k - t_{k-1}) \quad 0]^T$$
$$u_{k-1} = [\overline{V}_x \quad \overline{V}_y \quad 0]^T$$
$$w_{k-1} = [w_{x_{t_{k-1}}} \quad w_{y_{t_{k-1}}} \quad w_{z_{t_{k-1}}}]^T$$
(6)

Also, $E(e) = E(w) = 0$, $E(e, e^T) = R$ is a diagonal matrix whose entries are inversely proportional to the LOS observation variances and $E(w, w^T) = Q$ is also a diagonal matrix whose entries are inversely proportional to the GNSS variances. Assuming that $X_0 = 0$ and $E(X_0, X_0^T) = P_{0|0} = 0$, a

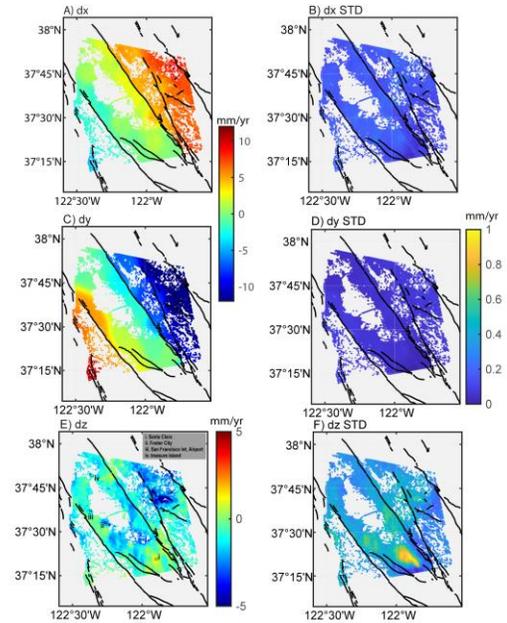

**FIGURE 3** Resolved 3D velocity field and associated standard deviation. A, B) $d_x$ and STD, C, D) $d_y$ and STD, E, F) $d_z$ and STD. Black lines are major fault lines.

### III. Experiment and Results

Figure 3 shows the obtained 3D velocities and associated standard deviations at the location of ~350,000 elite pixels obtained from applying the algorithm described above. Maps of $d_x$ and $d_y$ velocities are dominated by the motions along the major fault strands. In contrast, the $d_z$ map is mainly characterized by subsidence associated with compaction of sediments and landfills along the coasts, with rates up to -4 mm/yr affecting San Francisco International Airport, Foster City and Treasure Island, consistent with earlier studies [26, 29, 39, 53]. The standard deviation of the estimated 3D velocities also shows small overall values with medians of ~0.2, ~0.1 and ~0.4 mm/yr for $d_x$, $d_y$ and $d_z$ velocities, respectively. The STD map for $d_z$ show larger values in the south of the Bay over the Santa Clara valley, which is due to the non-linearity of the signal as discussed below.

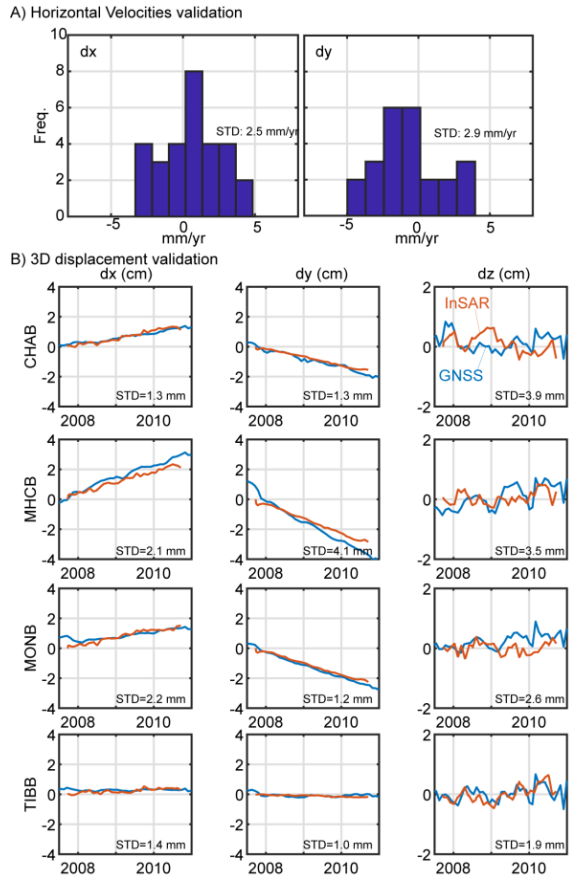

**FIGURE 4** Validating resolved 3D displacement field against GNSS check dataset horizontal velocities (A) and continuous GNSS time series (B). location of GNSS stations are shown in Figure 1. STD: standard deviation.

To validate the results, I use two datasets; firstly, I explore the GNSS check data points that were not used in the analysis. This dataset only includes horizontal velocities. To perform the validation test, I average the measurements of pixels within a 200 m distance of a given GNSS station. Figure 4A shows histograms comparing the obtained $d_x$ and $d_y$ velocities and that measured by the check stations. I find standard deviations of 2.5 mm/yr and 2.9 mm/yr for $d_x$ and $d_y$ velocities, respectively. Next, I compare the obtained 3D time series with that observed at the location of 5 continuous GNSS stations with measurements during the observation period that were not used in the analysis, whose locations are shown in Figure 1. I consider the LUTZ station as the reference point for both inversion results and the continuous GNSS network and thus, the time series in Figure 4B refers to the LUTZ station. I found an overall good agreement between datasets with maximum STD of differences of 2.2 mm/yr, 4.1 mm/yr and 3.9 mm/yr for $d_x$, $d_y$ and $d_z$, respectively.

## IV. Discussion and conclusions

Here I presented a framework based on Kalman Filter to combine InSAR 1D LOS time series and GNSS 2D velocities and solve for 3D displacement field time series. The approach is model-free and does not require any prior assumptions about the spatial and temporal behavior of the displacement field.

The experimental results show a good agreement between the 3D displacement field obtained from this approach and independent datasets, indicating the approach's success in resolving the evolution of complex displacement fields. Figure 5A shows examples of $d_z$ time series at San Francisco International Airport (SFO) and Santa Clara Valley, whose locations are marked in Figure 3E. The SFO time series was stable until mid-2009, when it experienced subsidence of up to 1 cm over 3 months, which was followed by an exponentially decaying subsidence, a characteristic behavior of compacting soil [54]. The time series of $d_z$ at Santa Clara Valley shows a sinusoidal behavior with 1 year period, which is characteristic of recharging and depleting aquifers, which I also reported in earlier studies [38, 55].

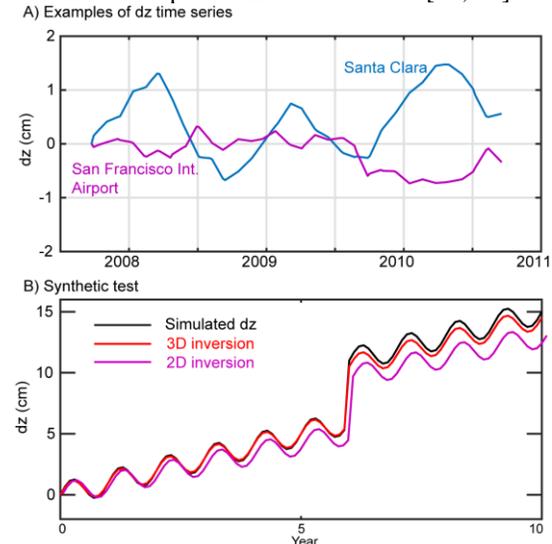

**FIGURE 5** A) example of $d_z$ time series at San Francisco International Airport and Santa Clara Valley, whose locations are shown in Figure 3E. B) Synthetic test comparing $d_z$ time series obtained from the method presented here (3D inversion) with that of Ozawa and Ueda [56] (2D inversion) that assumes $d_y$ displacement in north-south direction is zero.

The presented approach is applicable to an arbitrary set of overlapping SAR frames and GNSS tie points. The minimum number of SAR frames is 1, but the size of the GNSS tie dataset is case-dependent and varies, given the complexity of the horizontal velocity field. For the case of the San Francisco Bay Area, we found that a smaller set of GNSS stations can yield similar results since the majority of the horizontal displacement is caused by shearing along strike-slip faults [41].

The advantage of this approach over those that do not use GNSS tie points and only rely on the combination of ascending and descending frames (such as [56]) or prior information and models is demonstrated in Figure 5B. I simulated a time series comprising trend, annual variation, and a step and then created three interferometric datasets mimicking the experimental results shown in section IIC. Approaches similar to that of Ozawa and Ueda [56] express the slant-range changes by two components in east-west and up-down and assume the north-south component is zero. As seen, the approach presented here (red color) provides a better fit for simulated $d_z$ time series compared with that obtained following Ozawa and Ueda [56] approach (magenta color). This is because the assumption of zero $d_y$ displacement in the north-south direction, results in mapping the entire error caused by this assumption into the $d_z$ displacement. Similarly, the other approaches that apply prior model and information on behavior of 3D displacement field (e.g., [31, 33]) may under-perform when the assumed model is not accurate enough.

The presented approach in this article is particularly suitable for combining datasets from various SAR missions and GNSS observations to create seamless and continuous 3D observations of the displacement field. The obtained estimates of vertical land motion at mm-level precision and 10s-m spatial resolution are suitable for evaluating hazards due to natural and anthropogenic processes such as land subsidence [57] and relative sea-level rise [58].

## ACKNOWLEDGMENT

This research was supported by National Science Foundation (NSF) grants EAR-1357079 and EAR-0951430, and National Aeronautics and Space Administration (NASA) grants 16-ESI16-0021 and 80NSSC170567. The Alaska Satellite Facilities provided radar data. I thank UNAVCO for operating and maintaining the PBO GNSS stations and data archiving. These data and services are provided by the UNAVCO Facility with support from the NSF and NASA under NSF Cooperative Agreement No. EAR-0735156. Data for this study come from the Bay Area Regional Deformation Network (BARD), doi:10.7932/BARD, operated by the UC Berkeley Seismological Laboratory, which is archived at the Northern California Earthquake Data Center (NCEDC), doi: 10.7932/NCEDC.

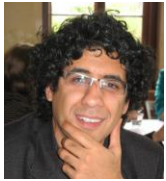

**Manoochehr Shirzaei** received a M.S. degree in surveying engineering from Tehran University, Tehran, Iran, in 2003 and a Ph.D. degree in geophysics and remote sensing from the University of Potsdam, Potsdam, Germany, in 2010. From 2011 to 2013, he was a Research scholar with the UC Berkeley seismology lab. Since 2020, he has been an Associate Professor of earth science and remote sensing with Virginia Polytechnic Institute and State University, VA, USA. He is the author of more than 70 articles. His research interests include SAR interferometry, machine learning, and crustal deformation processes. He is an Editor of the journal PNAS NEXUS.